\documentclass{article}

\usepackage{emulateapj,epsfig}

\slugcomment{Submitted to ApJ Letters}

\makeatletter

\newenvironment{inlinefigure}{%
\def\@captype{figure}%
\noindent\begin{minipage}{0.999\linewidth}\begin{center}}
{\end{center}\end{minipage}\smallskip}

\makeatother

\newcommand{\OVII}{O{\sc vii} }

\newcommand{\OVIII}{O{\sc viii} }

\newcommand{\OVIns}{O{\sc vi}}

\newcommand{\OVIIns}{O{\sc vii}}

\begin{document}

\title{The Temperature Structure of the Warm-Hot Intergalactic Medium}

\author{Naoki Yoshida\altaffilmark{1}, 
Steven R. Furlanetto\altaffilmark{2}, 
and Lars Hernquist\altaffilmark{3}}

\altaffiltext{1}{Department of Physics, Nagoya-University, Furocho, Chikusa, Nagoya 464-6082, Japan}
\altaffiltext{2}{Mail Code 130-33; California Institute of
  Technology; Pasadena, CA 91125}
\altaffiltext{3}{Harvard-Smithsonian Center for Astrophysics, 60 Garden Street, Cambridge, MA 02138}

\begin{abstract}

We study the temperature structure of the intergalactic medium (IGM)
using a large cosmological $N$-body/SPH simulation. We employ a
two-temperature model for the thermal evolution of the ionized gas, in
which we include explicitly the relaxation process between electrons
and ions. In the diffuse, hot IGM, the relaxation time is comparable
to the age of the Universe and hence the electron temperature in
post-shock regions remains significantly smaller than the ion
temperature.  We show that, at the present epoch, a large fraction of
the warm/hot intergalactic medium (WHIM) has a well-developed
two-temperature structure, with typical temperature differences of
order a few.  Consequently, the fraction of metals in various
ionization states such as O{\sc vi}, O{\sc vii}, and O{\sc viii}, as
well as their line emissivities, can differ locally by more than an
order of magnitude from those computed with a single-temperature
model,
especially in gas with $T \sim 10^7$ K.  It is thus necessary to
follow the evolution of the electron temperature explicitly to
determine absorption and emission by the WHIM.  Although equipartition
is nearly achieved in the denser intracluster medium (ICM), we find an
appreciable systematic deviation between the gas-mass weighted
electron temperature and the mean temperature even at half the virial
radii of clusters.  There is thus a reservoir of warm ($T_{e}<1$ keV)
gas in and around massive clusters. Our results imply that relaxation
processes need to be considered in describing and interpreting
observational data from existing X-ray telescopes as well as from
future missions designed to detect the WHIM, such as the {\sl Diffuse
Intergalactic Oxygen Surveyor} and the {\sl Missing Baryon Explorer}.

\end{abstract}


\section{Introduction}

The distribution of baryons in the Universe remains one of the
puzzling issues in modern cosmology.  At the present epoch, the total
amount of baryons inferred from a census of H{\sc i} absorption, gas
and stars in galaxies, and X-ray emission from hot gas in galaxy
clusters is far smaller than that predicted by nucleosynthesis
calculations (Persic \& Salucci 1992; Fukugita, Hogan \& Peebles 1998)
and by measurements of the cosmic microwave background radiation
(Spergel et al. 2003).  Hence, it is believed that a large fraction of
the baryons lies in an as yet undiscovered {\it dark} state.

Three dimensional hydrodynamic simulations of cosmic structure
formation (e.g., Cen \& Ostriker 1999; Dav\'e et al. 2001; Croft et al.
2002) predict that about 30--40\% of the baryons reside in the
so-called warm-hot intergalactic medium (WHIM) at the present epoch.
This gas is mostly shock-heated to a temperature of $\sim 10^5$--$10^7$ K
during large scale structure formation, and this relatively low
temperature makes its thermal emission difficult to detect with
conventional X-ray probes.

A variety of observational approaches have been considered for studying
the WHIM, including both ultraviolet (e.g., Furlanetto et al. 2003,
2004) and X-ray lines.  With respect to the latter possibility, there
have been several tentative claims that the WHIM has been detected
locally in absorption (e.g., Nicastro et al. 2002) and in emission
(Finoguenov et al. 2003).  Moreover, Yoshikawa et al. (2004) have
proposed using O{\sc vii}/O{\sc viii} lines to examine the WHIM in
detail with high spectral resolution X-ray detectors.  They argue that,
under a variety of assumptions, about half of the WHIM (by mass) can
be detected via oxygen line emission (see also Fang et al. 2004).
While the proposed missions appear promising, the detectability
depends crucially on the assumed metallicity and the relative
abundances of particular ions.

Previous theoretical studies used a single-temperature model for the
WHIM, under the assumption that equipartition is fully achieved.
However, the outskirts of galaxy clusters may actually have a
two-temperature structure (e.g., Fox \& Loeb 1997; Takizawa 1998;
Courty \& Alimi 2004).  Near galaxy clusters, infalling gas
shock-heats to roughly the virial temperature as kinetic energy is
converted into thermal energy.  Because the infalling ions carry
nearly all of the bulk kinetic energy (exceeding that contained in
electrons by a factor $\sim m_p/m_e$), the post-shock electrons can
gain thermal energy only via collisions with ions. If collisional
relaxation takes a time comparable to or longer than the age of the
Universe, the two temperatures, i.e. those of the electrons and ions,
remain different.  One can easily show that the relaxation time is
rather long for diffuse warm-hot gas around clusters and in
large-scale filamentary structures.  Since the relative abundance of
metal ions and their emissivities are sensitive to the electron
temperature, this has important implications for future observations
of the WHIM, particularly for the hotter component.

In this {\sl Letter}, we study the thermal evolution of the
intergalactic medium (IGM) using a large hydrodynamic 
\begin{figure*}[t]
\epsscale{2.0}
\plotone{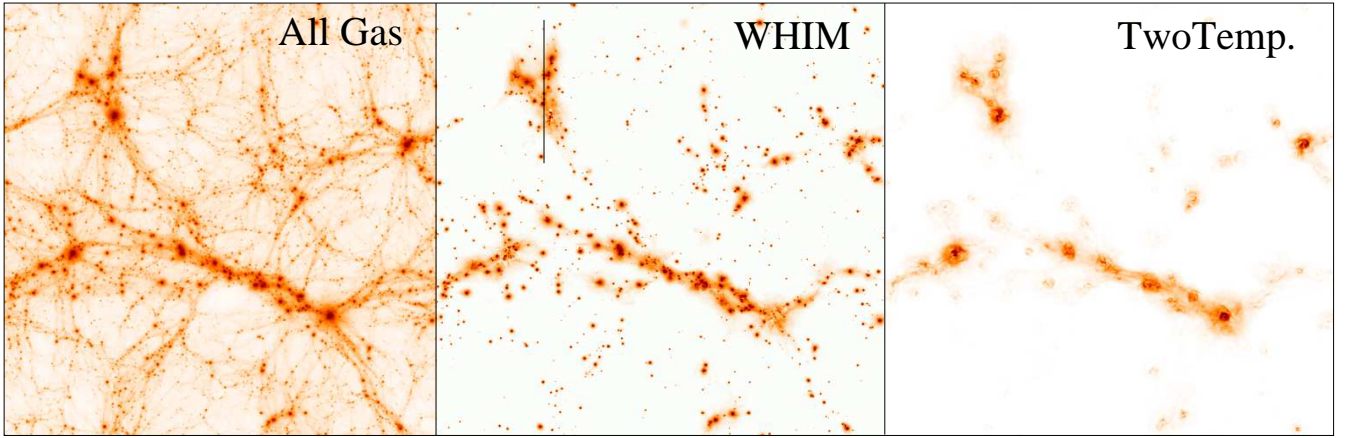}
\caption{The distribution of baryons (left), the warm/hot component
with $10^6 < T < 10^7$ K (middle), and gas with $T_{\rm e}< 0.5T_{\rm
i}$ (right) in a slab of $100\times 100\times 20$
($h^{-1}$Mpc)$^3$ at $z=0$. The vertical bar in the middle panel indicates a
`super-cluster' region, for which we compute the soft X-ray intensity
(see section 4 and Figure 5).
\label{plot1}}
\end{figure*}
\noindent simulation of
structure formation.  We show that a considerable fraction of the WHIM
indeed has a two-temperature structure, particularly in and around
rich clusters.  In these regions, the electron temperature is
typically smaller than the ion temperature by a factor of a few,
substantially modifying estimates of the abundances of ions and their
emissivities.

\section{The $N$-body/SPH simulations}

For our simulations, we use the parallel Tree-PM/SPH code GADGET2
(V. Springel, in preparation), which employs a fully conservative scheme for integrating
the equations of motion (Springel \& Hernquist 2002).
We implement non-equipartition processes between ions and electrons
following Fox \& Loeb (1997) and Takizawa (1998).
Equilibrium in an electron-proton plasma is achieved
in the following manner. After passing through a shock,
electrons and ions thermalize into (separate) Maxwellian distributions
on equilibration timescales $t_{\rm ee}, t_{\rm ii}$, with
$t_{\rm ii}/t_{\rm ee} \sim (m_{\rm i}/m_{\rm e})^{1/2} \sim 43$
(Spitzer 1962).  Here, $m_{\rm e}$ and $m_i$ are the
electron and ion masses and we have assumed protons dominate the
ionic component. Equipartition between protons and electrons is
achieved on an even longer timescale $t_{\rm ei} \sim (m_{\rm i}/m_{\rm e})t_{\rm ee}$.  
Hence, we assume that electrons and ions quickly achieve
Maxwellian distributions with temperatures $T_{\rm e}$ and $T_{\rm i}$,
respectively, and consider only the non-equipartition effect.
Spitzer (1962) showed that the appropriate timescale is given by
\begin{equation}
t_{\rm ei}=\frac{3m_{\rm e}m_{\rm i}}{8(2\pi)^{1/2}n_{i}Z_{i}^{2}
e^4\ln \Lambda}
\left(\frac{k T_{\rm e}}{m_{\rm e}}+\frac{k T_{i}}{m_{i}}\right)^{3/2},
\end{equation}
where $e$ is the electric charge, $Z_{\rm i}$ is the charge of an ion,
$n_{\rm i}$ is the ion number density, and $\ln \Lambda$ is the
Coulomb logarithm, which is given by
\begin{equation}
\ln \Lambda = 37.8 + \ln \left(\frac{T_{e}}{10^{7} {\rm K}}\right) -
\frac{1}{2}\ln \left(\frac{n}{10^{-5}{\rm cm}^{-3}}\right).
\end{equation}
For a hot, fully ionized plasma,
\begin{equation}
t_{\rm ei}=6.3\times 10^{8} {\rm yr}\;
\frac{(T_{\rm e}/10^{7} {\rm K})^{3/2}}
{(n/10^{-5}{\rm cm}^{-3})(\ln \Lambda/40)},
\end{equation}
and this timescale can be comparable to the Hubble time in regions
with $T \ga 10^7$ K and overdensities $10-100$ with respect to the
cosmic mean.
The evolution of the electron temperature is given by
\begin{equation}
\frac{dT_{\rm e}}{dt}=\frac{T_{\rm i}-T_{\rm e}}{t_{\rm ei}}
+(\gamma-1)\frac{T_{\rm e}}{n}\frac{dn}{dt}.
\end{equation}
%moved from section 3
We do not consider heat conduction on the assumption that it will be
suppressed by tangled magnetic fields.  Some other electron heating
processes driven by plasma instabilities have also been proposed
(Laming 2000), but detailed calculations
in the context of supernova remnants
suggest that these mechanisms convert less than $20$\% of the bulk
kinetic energy to electron thermal energy (e.g., Cargill \&
Papadopoulos 1988), so we neglect these effects.  Our treatment
therefore maximizes the possible offset between $T_{\rm e}$ and
$T_{\rm i}$.

We work with a flat $\Lambda$-dominated cold dark matter cosmology
with matter density $\Omega_{0}=0.3$, cosmological constant
$\Omega_{\Lambda}=0.7$ and expansion rate at the present time
$H_{0}=70$ km s$^{-1}$ Mpc$^{-1}$. We set the baryon density
$\Omega_{\rm b}=0.04$ and the normalization parameter $\sigma_8
=0.9$. Our simulation employs $324^3$ cold dark matter particles and
the same number of non-radiative gas particles in a cosmological
volume 100 $h^{-1}$ Mpc on a side.
The mass per gas particle is
$3.26\times 10^8 h^{-1}M_{\odot}$, and the nominal gas mass resolution
is about $10^{10}h^{-1}M_{\odot}$ for our choice of the number of SPH
neighbors, $N=32$.  We note that the CDM initial conditions and the
large simulation volume allow us to correctly model the merging
process during the formation of filamentary structures and
the assembly of massive halos which contribute to heating of the
IGM as well as the intracluster medium (ICM; e.g., Keshet et al. 2003).

\section{Results}

Figure 1 shows the baryon distribution at $z=0$ in a slice 20
$h^{-1}$ Mpc thick (top),
the hot component with temperature
$10^6 < T < 10^7$ K (middle), and gas in which
$T_{\rm e} < 0.5 T_{\rm i}$.
The latter two components are quite similar,
suggesting that much of the hotter WHIM has a two-temperature structure.
The distribution of the two-temperature gas appears quite complex even
in clusters, 
\begin{inlinefigure}
\hspace{-1cm}\resizebox{7.5cm}{!}{\includegraphics{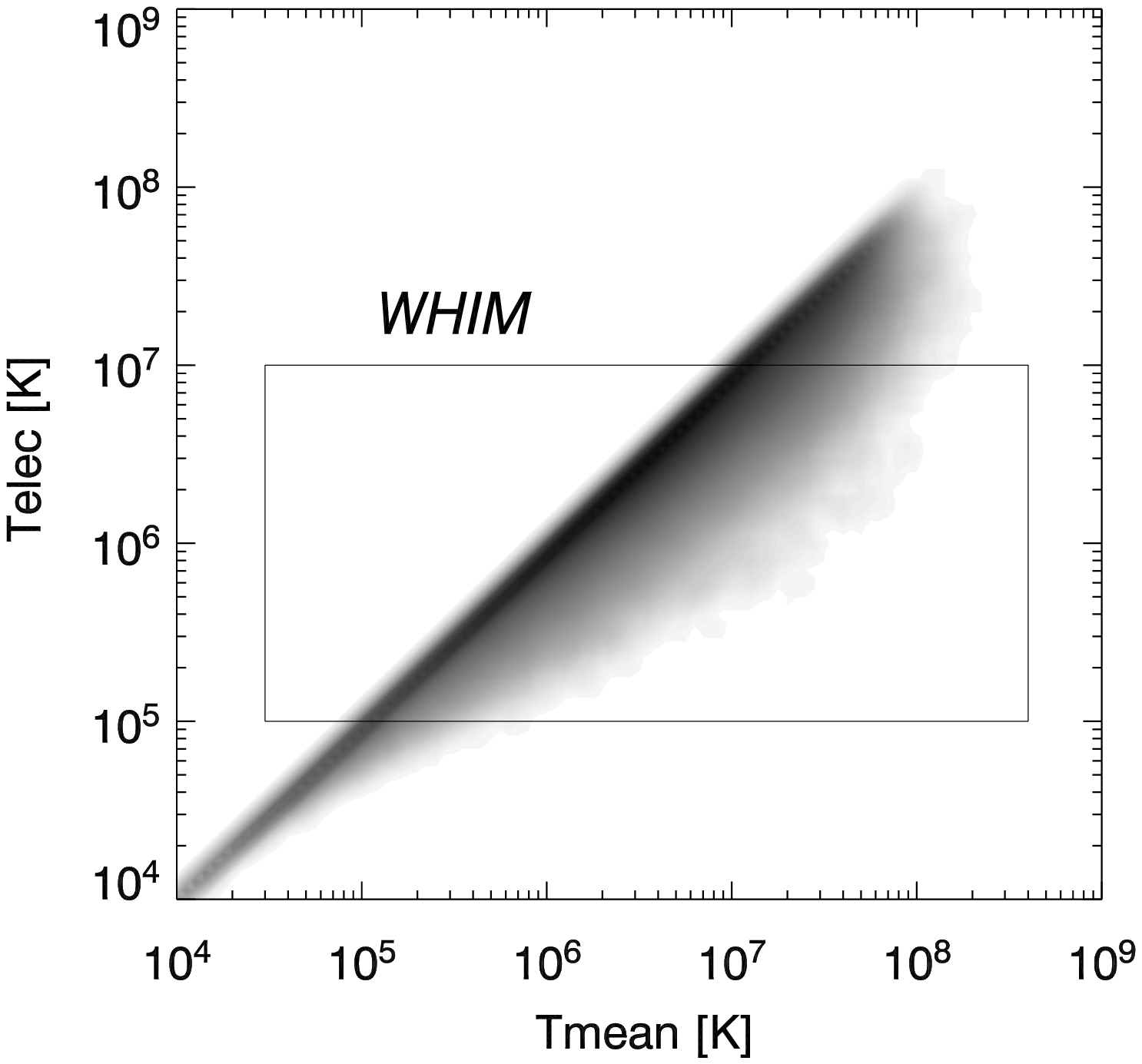}}
\caption{Electron temperature plotted against
the mean gas temperature. 
The number density of gas particles is shown in grey-scale.
The rectangular box indicates the so-called
WHIM which we define to have $10^5 < T_{\rm e} < 10^7$K.
\label{plot2}}
\end{inlinefigure}

\noindent indicating that shocks induced by mergers during
hierarchical assembly propagate and leave two-temperature gas within
clusters.  We have checked the time variation of the specific entropy
of the gas elements and verified that the two-temperature regions have
recently been shock-heated.
%revise
In the gas around large-scale structure, {\it continuous}
shock-heating disturbs equipartition
sufficiently close to the shock, while more distant gas is often
relaxed.

Figure 2 shows the electron temperature $T_{\rm e}$ plotted against
the mean temperature 
$\bar{T}=(n_{\rm e}T _{\rm e}+n_{\rm i}T_{\rm i})/(n_{\rm e}+n_{\rm i})$, 
which is the temperature we would obtain
in a single temperature model.  Interestingly, the two-temperature
deviation is largest in the WHIM regime (indicated by the rectangular
box).  Note that there are some gas elements with $T_{\rm e} \ll
T_{\rm i}$. This is because our simulation does not include
electron heating processes other than Coulomb collisions.

Figure 3 shows the radial distribution of $T_{\rm e}$ (solid line) and
$\bar{T}$ (dashed line) for one of the most massive halos in our
simulation.  We used a friends-of-friends group-finder to locate dark
matter halos and define the virial radius such that the mean matter
density within this radius is 200 times the critical density.  The
halo we consider here has a virial mass of $3\times 10^{14} h^{-1}
{\rm M}_{\odot}$ and a virial radius of $1.1 h^{-1}$ Mpc.  In Figure
3, the difference between the two temperatures is large at $r > 1
h^{-1}$ Mpc and appreciable even at $r < 1 h^{-1}$ Mpc.  The overall
profile looks similar to those found in simulations of single clusters
by Takizawa (1998) and Chieze et al. (1998).  Owing to the difference
in relaxation time (see eq. [3]) in the dense central part and in the
low density outer regions, the electron temperature profile appears
steeper than that of the mean temperature at $r > 0.1 h^{-1}$ Mpc.
The bottom panel of Figure 3 shows the baryonic mass fraction as a
function of both $T_{\rm e}$ and $\bar{T}$ (thick and thin histograms,
respectively).  We selected gas particles within 5 $h^{-1}$ Mpc of the
cluster center.  The distribution of $T_e$ is skewed significantly to
smaller values. The total gas mass contained in this region ($r < 5
h^{-1}$ Mpc) is $1.0\times 10^{14} h^{-1} \ {\rm M}_{\odot}$, and
nearly 50\% of it has $T_{\rm e}<1$ keV; i.e., the mass of the warm

\begin{inlinefigure}
\vspace{-5mm}
\resizebox{9cm}{!}{\includegraphics{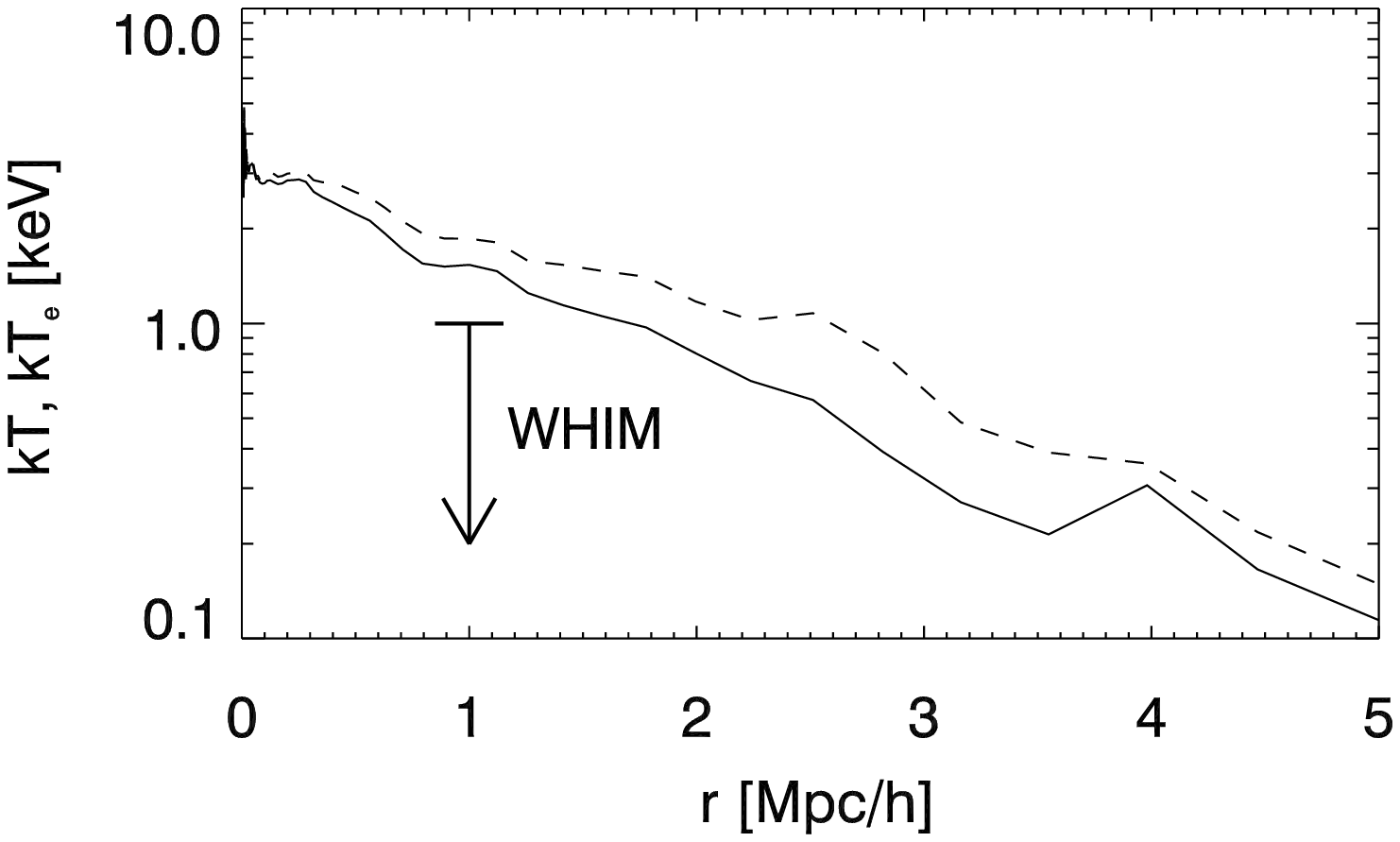}}\vspace{-0.7cm}
\resizebox{9cm}{!}{\includegraphics{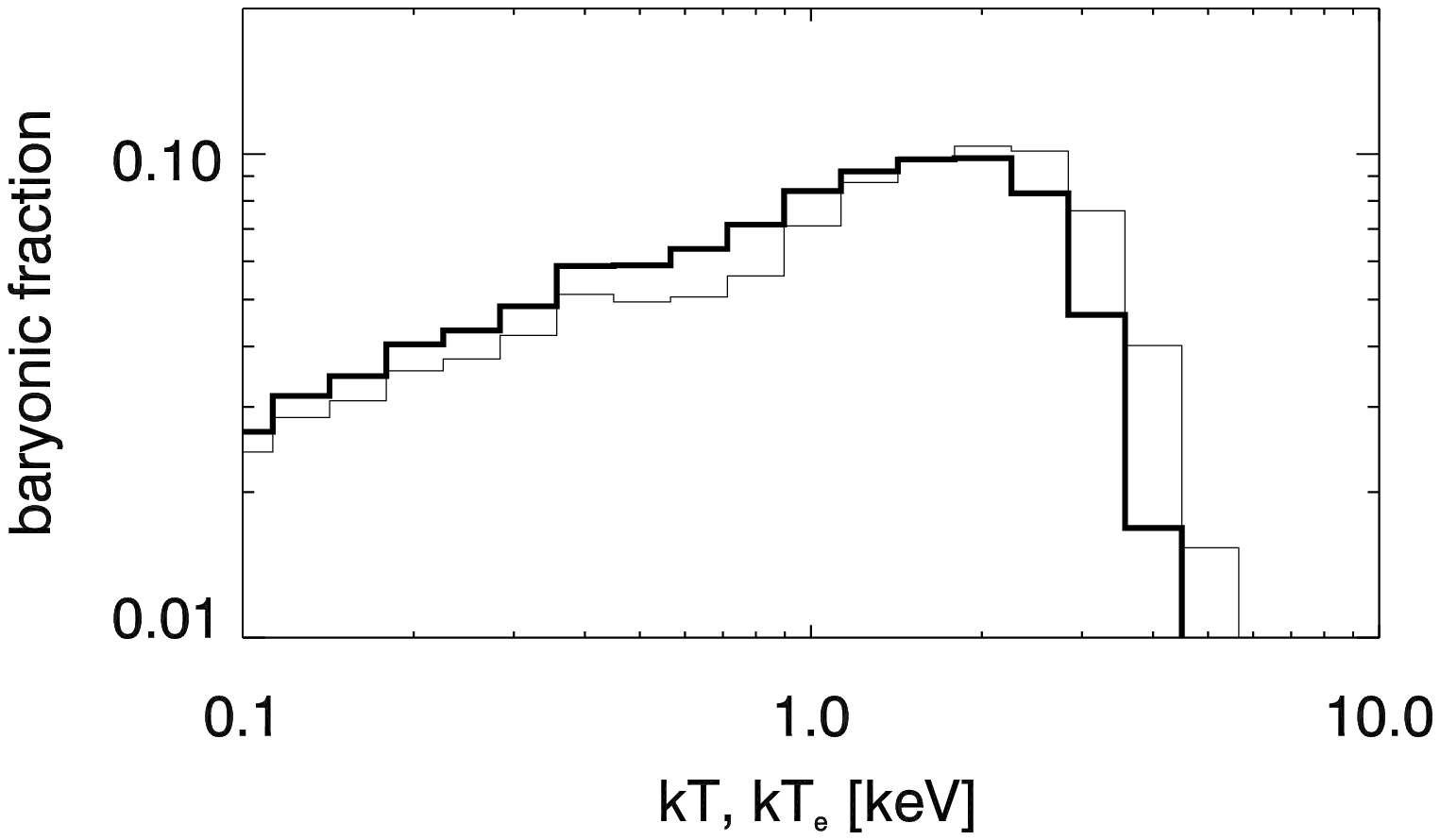}}
\caption{Radial distribution of the electron temperature
(solid line) and the mean temperature (dashed line) in one of the
most massive clusters in our simulation (top panel). 
Bottom panel shows the mass fraction of gas within 5$h^{-1}$Mpc 
from the cluster center in temperature bins,
using the electron temperature (thick histograms) and the mean 
temperature (thin histograms).
\label{plot3}}
\end{inlinefigure}

\noindent component surrounding the cluster is greater than the gas mass 
within
the cluster itself.

\section{Cosmological Implications}
The two-temperature structure of the WHIM and ICM has many important
implications.  For example, the relative fractions of metal ions are
of considerable interest for WHIM observations.  Figure 4 shows the
fraction of \OVIns, \OVIIns, and \OVIII ions as a function of
temperature assuming collisional ionization equilibrium.
\footnote{The assumption of collisional ionization equilibrium may not
be strictly valid in the WHIM.  We nevertheless adopt this
commonly-made assumption to compute the line emissivities in a
post-processing manner.}  Clearly the abundances have a strong
dependence on {\it electron} temperature, because the populations are
primarily determined by electron collisions (photo-ionization is
unimportant for these high-level ions unless $T \ll 10^{5.5}$K or
the incident X-ray intensity is high).  The bottom panel indicates the
possible errors caused by assuming a single temperature for the WHIM.
A factor of two systematic shift in temperature, typical of the
offsets we find between $\bar{T}$ and $T_{\rm e}$
near shocks,
can lead to significant over/under-estimates of the abundances.

We further quantify the importance of non-equipartition with a model
oxygen line emission map.  We put the super-cluster region indicated
in Figure 1 at $z=0.03$ and compute the surface brightness of \OVII
emission following the procedure of Yoshikawa et al. (2004).  We adopt
a simple relation between the gas metallicity and the local gas
density of the form $Z=0.02 (\rho_{\rm gas}/ \bar{\rho})^{0.3} \ {\rm Z}_{\odot}$, 
where $\bar{\rho}$ is the mean baryon density.  We then
compute the surface brightness for both $T_{\rm e}$ and $\bar{T}$.
Figure 5 shows the 

\begin{inlinefigure}
\resizebox{7.5cm}{!}{\includegraphics{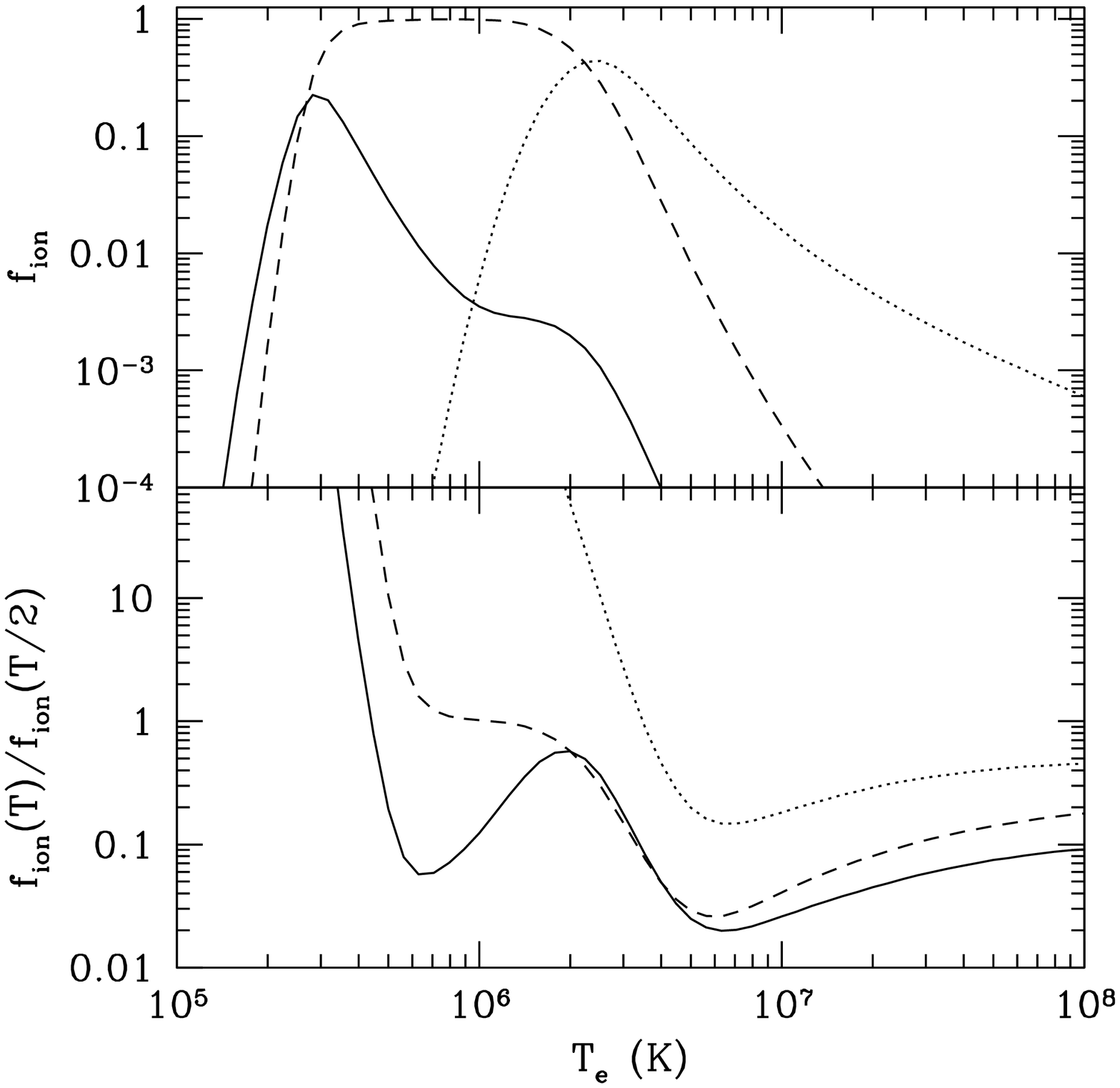}}
\caption{Relative abundance of oxygen ions $f_{\rm O_{\rm VI}}$ (solid),
$f_{\rm O_{\sc VII}}$ (dashed), and $f_{\rm O_{\rm VIII}}$ (dotted) as a function of
electron temperature.  The bottom panel shows the ratio $f(T)/f(T/2)$
which indicates the error induced by a factor of two shift
(misestimate) in $T_e$.
\label{plot4}}
\end{inlinefigure}

\noindent derived profiles.  While the overall results appear
similar in the two cases, there are substantial local deviations.  In
particular, around some bright spots, the emissivity is greater by up
to an order of magnitude when computed using the electron temperature.
The ratio of the surface brightness $S(T_{\rm e})/S(\bar{T})$ is shown in
the bottom panel of Figure 5.  There, we mark the regions with
$S(T_{\rm e})> 10^{-11} \ {\rm erg s}^{-1} \ {\rm cm}^{-2} \ {\rm
sr}^{-1}$, which is the nominal detection limit of the proposed DIOS
mission (Yoshikawa et al. 2003).

Similar considerations may also apply to efforts to study the WHIM
using UV observations (e.g., Furlanetto et al. 2004),
although the steep temperature dependence of $t_{\rm ei}$ implies that
equipartition is more accurate in this regime (see Fig. 2).
In principle, one can measure the degree of equilibration using
various line intensities from far UV to soft X-rays (e.g., Ghavamian
et al. 2001; Raymond et al. 2003) as well as the metal line Doppler
widths.  Measuring the age of the WHIM from the degree of
equilibration will put a stringent constraint on theoretical models of
large-scale {\it baryonic} structure formation.

\begin{inlinefigure}
\vspace*{-5mm}
\resizebox{8.5cm}{!}{\includegraphics{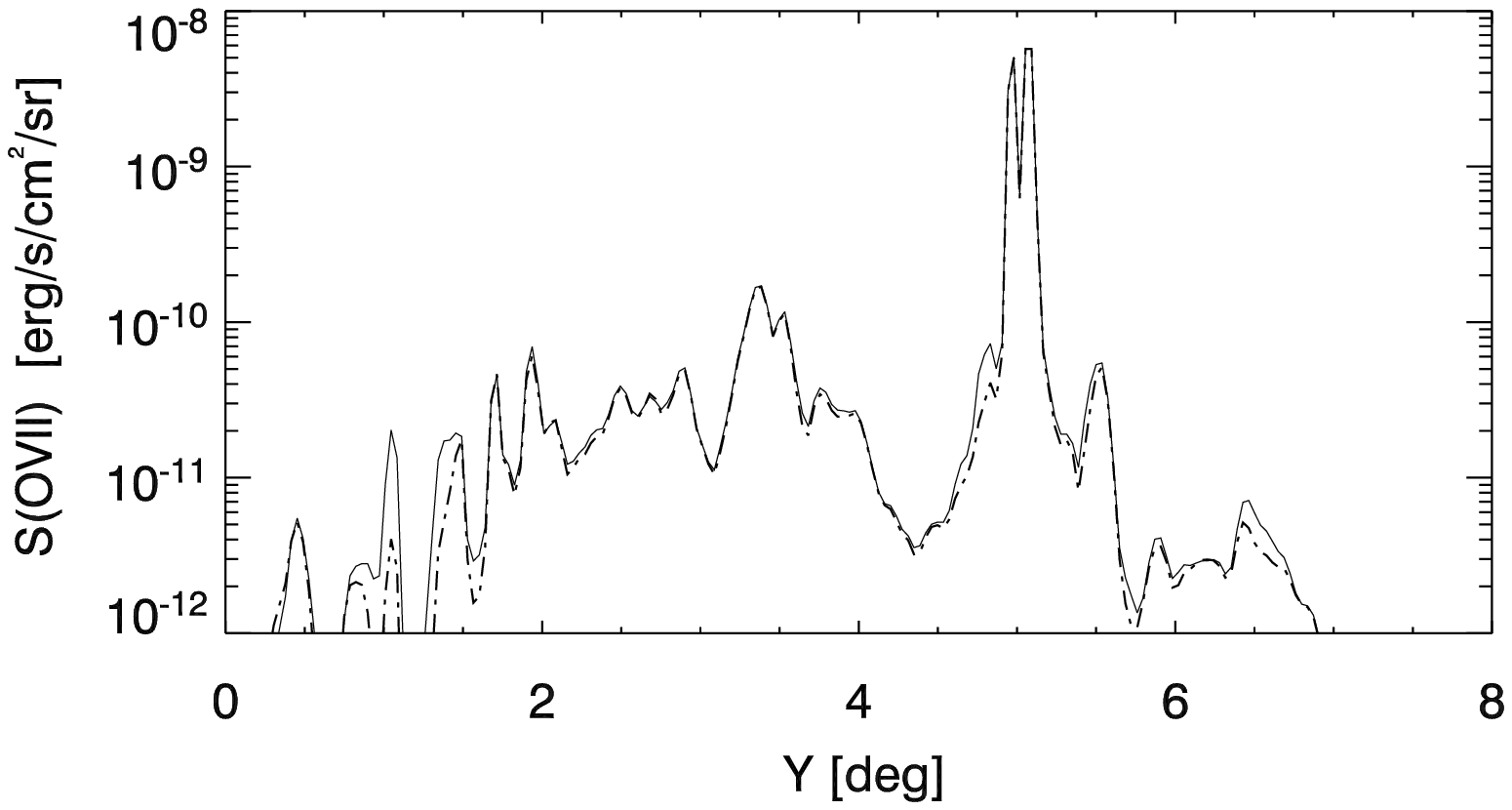}}
\vspace{-0.5cm}
\hspace{-1cm}\resizebox{8.5cm}{!}{\includegraphics{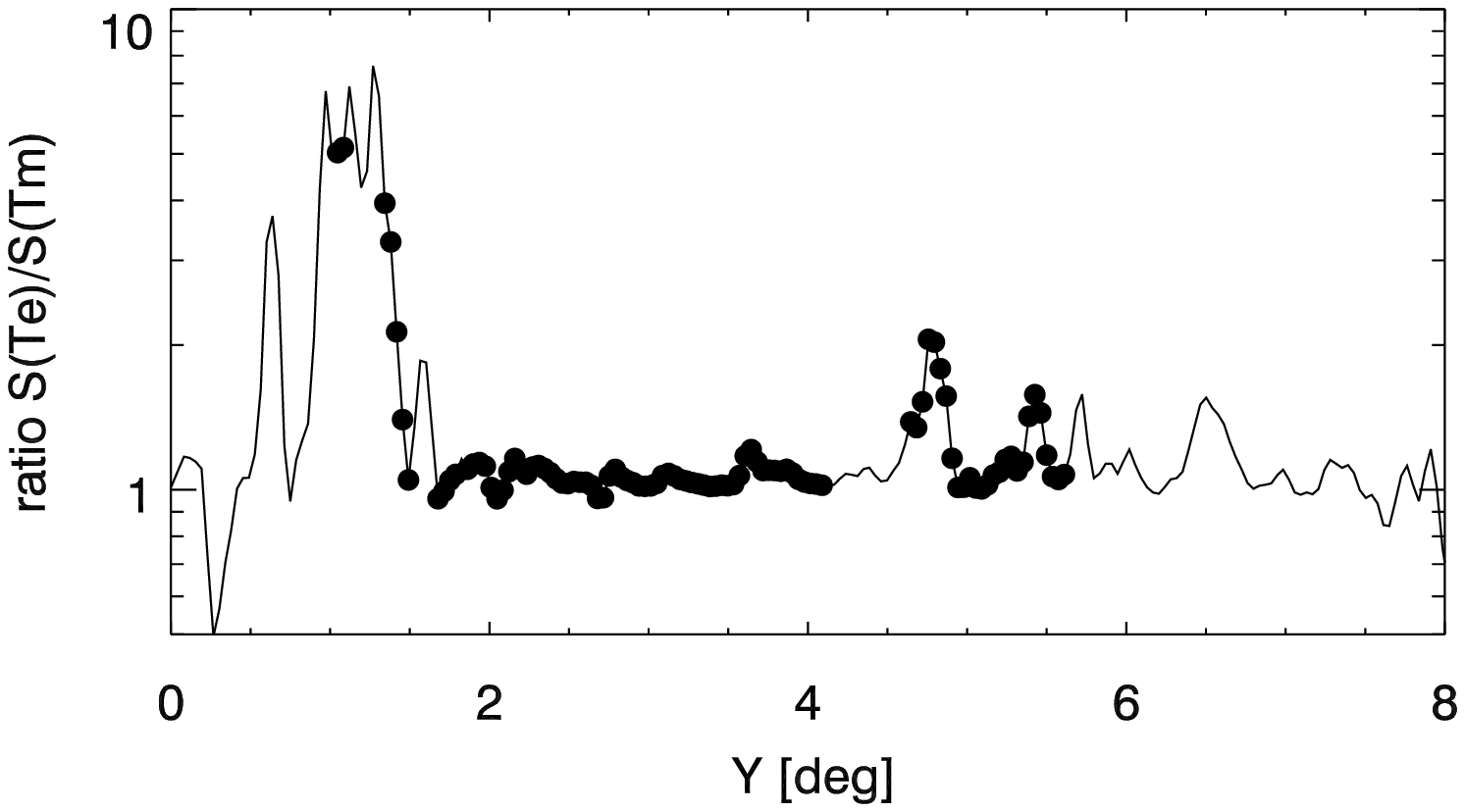}}
\caption{Surface brightness of \OVII emission for the region indicated
in the middle panel of Figure 1 using $T_{e}$ (solid line) and
$\bar{T}$ (dot-dashed line). Bottom panel shows the ratio
$S(T_{e})/S(\bar{T})$. Dots in the bottom panel indicate the region
where $S(T_{e})$ is greater than $10^{-11} \ {\rm erg s}^{-1} \ {\rm
cm}^{-2} \ {\rm sr}^{-1}$.
\label{plot5}}
\end{inlinefigure}

Although our simulation does not include radiative cooling, we note
that the lower electron temperature can have a significant impact on
the radiative cooling efficiency in dense regions. 
The gas cooling rate has a very steep temperature dependence 
at $10^5< T <10^6$K regardless of
the metallicity (Sutherland \& Dopita 1993).  As shown in Figure 2,
recently shock-heated gas with $\bar{T}\sim 10^6$K may have an
electron temperature $T_{\rm e}\sim 10^5$K. Such gas should cool
rapidly and in a highly non-equilibrium manner.  This warrants further
study of the effect of the two-temperature structure on the IGM using
a simulation with radiative cooling.

Overall, our simulation suggests that theoretical studies of the WHIM
and correct interpretation of the observational data from current and
future X-ray missions requires explicit consideration of these
relaxation processes.

We thank Kohji Yoshikawa for providing emissivity data.  The
simulations were performed at the Center for Parallel Astrophysical
Computing at the Harvard-Smithsonian Center for Astrophysics.

\end{document}